\documentclass[prb,a4paper,twocolumn]{revtex4}

\usepackage{graphicx}
\usepackage{subfigure}
\usepackage{color}
\usepackage{amsmath}
\begin{document}

\title  {Unusual Low-Energy Collective Charge Excitations in High-$T_c$ Cuprate Superconductors}

\author{Vyacheslav M. Silkin}
\affiliation{Donostia International Physics Center (DIPC), 20018 San Sebasti\'an/Donostia, Basque Country, Spain}
\affiliation{Departamento de Pol\'{\i}meros y Materiales Avanzados: F\'{\i}sica, Qu\'{\i}mica y Tecnolog\'{\i}a, Facultad de Ciencias Qu\'{\i}micas,
Universidad del Pa\'{\i}s Vasco UPV/EHU, 20080 San Sebasti\'an/Donostia, Basque Country, Spain}
\affiliation{IKERBASQUE, Basque Foundation for Science, 48013 Bilbao, Basque Country, Spain}

\author{Stefan-Ludwig Drechsler}
\affiliation{Leibniz Institute for Solid State and Materials Research IFW Dresden,
Helmholtzstr. 20, 01069 Dresden, Germany}

\author{Dmitry V. Efremov}
\affiliation{Leibniz Institute for Solid State and Materials Research IFW Dresden,
Helmholtzstr. 20, 01069 Dresden, Germany}

\begin{abstract}
 Despite decades of intensive experimental and theoretical efforts, the physics of cuprate high-temperature superconductors in general, and, in particular, their normal state, is still under debate. Here, we report our investigation of low-energy charge excitations in the normal state. We find that the peculiarities of the electronic band structure at low energies have a profound impact on the nature of the intraband collective modes. It gives rise to a new kind of mode with huge intensity and non-Lorentzian spectral function in addition to well-known collective excitations like conventional plasmons and spin fluctuation.
We predict two such modes with maximal spectral weight in the nodal and antinodal directions. Additionally, we found a long-living quasi-one-dimensional plasmon becoming an intense soft mode over an extended momentum range along the antinodal direction. These modes might explain some of the resonant inelastic X-ray scattering spectroscopy data.

\end{abstract}


\maketitle

Collective electronic excitations, plasmons, in cuprates attracted considerable attention soon after the discovery  of high-temperature superconductivity.\cite{bemuzpb86}
The early electron energy-loss spectroscopy (EELS) \cite{nuroprb89,ronuzpb90,nuecprb91,grpaprb99,rokojesrp14} measurements reported its existence and dispersion at energies around 1 eV. This issue was also discussed theoretically.\cite{ruprb87,grprb88,krmoprb88,bimoprb03,mahaprb08}

The revival of interest in cuprate high-temperature superconductors (HTSC) in recent years has been driven by advances
in resonant inelastic X-ray scattering spectroscopy (RIXS).\cite{amvarmp11} This allows the direct observation of collective modes and
excitations in both electron-doped \cite{hechn18,liyuqm20,heboprb23} and hole-doped \cite{nazhprl20,sihuprb22} cuprates.
Most of the charge excitations and collective modes have been successfully identified with the help of theoretically calculated incoherent
electron-hole (e-h) excitations, damped magnetic collective modes, and conventional plasmons.  However, some of the observed collective modes nevertheless exhibit
many unusual features, and the full picture is still not yet fully understood.


Since the electronic structure in cuprates has a quasi-two-dimensional character, plasmons also exhibit
a quasi-two-dimensional behavior.
This is primarily reflected in their
gapless dispersion $\omega_p(\mathbf{q}) \propto q^{1/2}$ at small in-plane momenta ${\bf q}$, in
contrast to the
dispersion with a large gap of the order of a few eV by 3D plasmons.\cite{stprl67}  This mode is further transformed into  a
gapless plasmon band in  quasi-2D layered materials, because of the interlayer long-range Coulomb interaction.\cite{feap74,daquprb82}
Incoherent e-h pairs represent another kind
of electronic excitations, which  forms a structureless continuum in 2D- and 3D-systems.
This types of excitations are believed to be dominant in cuprates,\cite{krmoprb88,hechn18}
resulting in strongly overdamped  charge excitations at low energies.

In this paper, we focus on the low energy-momentum phase space in cuprates.
By calculating the dielectric  and  loss functions with the use of
a realistic dispersion of the Bi-2212 partly filled energy band, we found a number of new collective charge excitations in the form and origin rather
different from conventional plasmons. They clearly
exist in the low-energy region due to their extraordinary total large oscillator	strength in spite of sizably strong detrimental,
 incoherent intraband e-h
pairs leading to visible remnants of charge excitations apart from the broad maximum of the loss function.


In  the random-phase approximation (RPA) used here the complex
dielectric function at momentum transfer ${\bf q}$ and  energy  transfer $\omega$ can be found as
\begin{equation}
 \label{eps}
 \epsilon({\bf q},\omega) \equiv \epsilon_1({\bf q},\omega)+{\rm i}\epsilon_2({\bf q},\omega) = \epsilon_{\infty}
- V({\bf q})\chi^o({\bf q},\omega),
\end{equation}
where $\epsilon_1({\bf q},\omega)$ and  $\epsilon_2({\bf q},\omega)$ are the real and imaginary parts of  $\epsilon({\bf q},\omega)$,
respectively. Notice, that in this work the momenta are expressed in units of the reciprocal lattice parameter $a$ ($1/a$).
In Eq.\ (\ref{eps}) $V({\bf q})$ denotes
the  Coulomb interaction  and $\chi^o({\bf q},\omega)$ is the density-density response function of non-interacting
electrons. Below we use the Coulomb interaction in the
two-dimensional form $V({\bf q})=2\pi/q$, but our results are robust with respect to the choice of the electron-electron interaction. We expect that similar results should be obtained with any form of $V({\bf q})$ in the metallic state away from metal-insulator transition.\cite{mahaprb08}
The
background dielectric constant  $\epsilon_{\infty}$ for Bi-2212  is $\epsilon_{\infty}$=4.5.\cite{letrprx16}
Since  at the ${\bf q}$'s of interest here, the second part on the right-hand side of Eq. (\ref{eps}) is significantly larger than
 $\epsilon_\infty$	the results presented here are insensitive to its exact value.
Therefore we employ $\epsilon_{\infty}$=1 for our calculations, i.e.,\ consider a freestanding Bi-2212 monolayer case.

The density-density response function $\chi^o({\bf q},\omega)$ is obtained using the conventional expression
\begin{equation}
\chi^o({\bf q},\omega) = \frac{2}{S}\sum_{\bf k}^{BZ}\frac{f_{\bf k}-f_{{\bf k}+{\bf q}}}{\varepsilon_{\bf k}-\varepsilon_{{\bf k}+{\bf
q}}+\omega+{\rm i}\eta}.
\label{chi_0}
\end{equation}
Here the factor 2 accounts for the two spin orientations, $S$ is the unit cell size, $f_{\bf k}$ is the Fermi distribution function at temperature $T=0$,
  $\varepsilon_{\bf k}$ is the dispersion of the quasiparticles, and $\eta>0$ is an infinitesimal damping parameter.
 For simplicity, we set the transition probability matrix elements equal to one, which is a good approximation at small and moderate momenta.\cite{jomaprb06,faarprb12}
In order  to avoid  an additional	
"artificial damping" due to using a finite value of $\eta$, in  our
numerical calculations, we found
the imaginary part $\chi^o(\mathbf{q},\omega)$ through the spectral function.
 Then the real part of $\chi^o(\mathbf{q},\omega)$ is restored using the Kramers-Kronig relation. For the calculations we use a mesh of
 2500$\times$2500 ${\bf k}$-points when summed over the Brillouin zone (BZ).

\begin{figure}
\centering
\includegraphics[width=0.49 \textwidth]{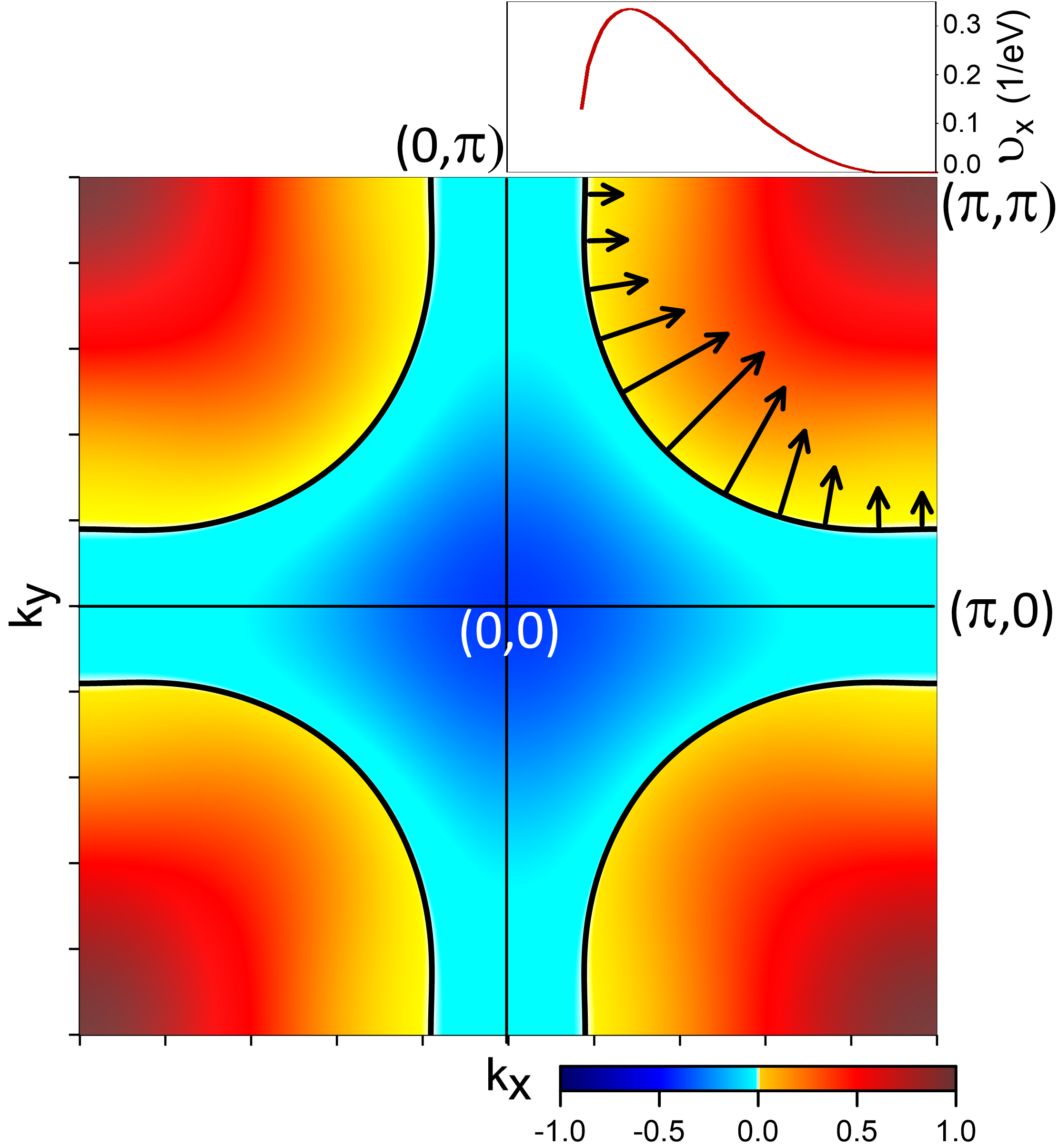}
\caption{2D plot of the empirical energy dispersion of the quasiparticles for the Bi-2212 tight-binding band structure.\cite{noraprb95}
		The Fermi level is marked by black solid lines. In the top-right quadrant, the arrows indicate the direction and the relative
magnitude of the quasiparticle velocities at the Fermi level.
On the top, the Fermi velocity component $\upsilon_x$ versus $k_x$. Notice that close to the antinodal direction in the interval of $k_x$
between $\approx 0.8\pi$ and $\pi$ it is almost zero.
}
\label{2D-band}
\end{figure}

For the dispersion of the quasiparticle energies $\varepsilon_{\bf k}$, shown in the density  plot in Fig.\ \ref{2D-band},
we use the empirical extended one-band tight-binding (TB) model for Bi-2212 proposed by Norman {\it et al.}\cite{noraprb95}
It reads
\begin{equation}\label{eq.TB}
	\varepsilon(\mathbf{k}) = \sum_i c_i \eta_i (\mathbf{k})
\end{equation}
with coefficients $c_i$ and functions $\eta_i(\mathbf{k})$ given in Tab. \ref{tab.parameters_of_TB}.  This TB band structure corresponds to a 17.6$\%$ hole doping.

\begin{table}
\begin{center}
\begin{tabular}{|c|c|}
	\hline
	$c_i$ (in eV)&  $\eta_i(\mathbf{k})$  \\
	\hline
	0.1305&  1 \\
	-0.5951& $\frac{1}{2}(\cos k_x + \cos k_y)$ \\
	0.1636 & $\cos k_x  \cos k_y $  \\
	-0.0519& $\frac{1}{2}(\cos 2k_x + \cos 2k_y) $ \\
	-0.1117 &$ \frac{1}{2}(\cos2 k_x \cos k_y + \cos k_x \cos 2 k_y)$ \\
	0.0510 &  $\cos2 k_x  \cos 2k_y $\\
	\hline
\end{tabular}
\end{center}
	\caption{ Parameters of the tight-binding model for Bi-2212.}
\label{tab.parameters_of_TB}
\end{table}

Having obtained  $\epsilon({\bf q},\omega)$, we calculate the electron energy-loss function, which is determined by the imaginary part of the
inverse dielectric function  $-\mbox{Im}[\epsilon^{-1}({\bf q},\omega)]$.
The fluctuation-dissipation theorem relates it to the dynamic structure factor $S({\bf q},\omega)$, which contains the complete information
about the density-density correlations and their dynamics.\cite{vapr54}
Reasonably sharp peaks in the measured spectra are usually regarded as
a signature of collective electronic excitations in the system (plasmons). The frequently used
theoretical definition of a plasmon with energy $\omega_p$ in textbooks is $\epsilon({\bf q},\omega=\omega_p)=0$, i.e.,\ the real and
imaginary parts of $\epsilon$ must be zero at $\omega=\omega_p$. However, in
real materials such a
condition is never fulfilled due to unavoidable losses. Therefore a more general
definition is used for the existence of a plasmon
resulting in a peak in the loss function. Then the practically
 used conditions are (a) a distinct peak in the loss function centered at $\omega_p$, (b) the presence of a zero crossing in the real part of $\epsilon$ at sufficiently the same energy,
and (c) a local minimum in the imaginary part $\epsilon_2$ in the close energy region.

For comparison, we realized similar calculations for the one- and two-component free-electron gas models. The respective parameters - Fermi wave
vectors and Fermi velocities - were adjusted to reproduce the energy positions of the peaks in $\epsilon_2$ and the 2D plasmon energies
obtained with the TB band structure.

The data presented below were obtained with the use of the 2D Coulomb potential. For comparison, we recalculated the data with the
3D potential in Eq.\ (\ref{eps}) and found out that regarding
the momenta where the modes found here appear, the type of the potential $V({\bf q})$ has only a negligible effect on the excitation spectra. This is because the shape of
the dielectric function and the loss function in the energy interval of interest here is essentially independent of $V(\bf q)$ since the
amplitude of the parts in $\epsilon$ at small momenta is large in comparison to $\epsilon_{\infty}$ (4.5 in Bi-2212).
Only approaching the BZ boundaries, the data become visibly dependent on the $V({\bf q})$  shape.
This is completely distinct from the behavior of conventional 2D- and 3D-plasmons.

There are several investigations in the literature concerning the interplay of plasmons and correlation
effects (see e.g., van Loon et al.\cite{vahaprl14}) in materials close to metal-insulator transition, where results obtained by the RPA have
been questioned. The authors compare their results with the free-electron case treated within the RPA and stress a reduction of the dispersion caused by the Hubbard U. In our approach a clear band renormalization as compared to density functional theory (DFT) band structure calculations by a factor of about three is phenomenologically taken into account using a generalized tight-binding approximation derived from experimental angle-resolved photoemission spectroscopy  (ARPES) data. Other issues like the claimed tendency to localization in cuprates are beyond of the scope of the present paper. Anyhow, they are obviously not relevant  for the nearly optimal doping regime and the region of relative small transferred  momenta considered here.
The possibility to describe successfully within the RPA the anisotropic plasmon dispersion measured by Nücker et al.\cite{nuecprb91} at high energies up to relatively large wave vectors has been demonstrated in the work by Grigoryan et al.\cite{grpaprb99} Based on these arguments, we believe that in a phenomenological RPA-based approach explicit correlation effects might be ignored at least at first approximation.

\begin{figure*}
\centering
\includegraphics[width=0.49\linewidth]{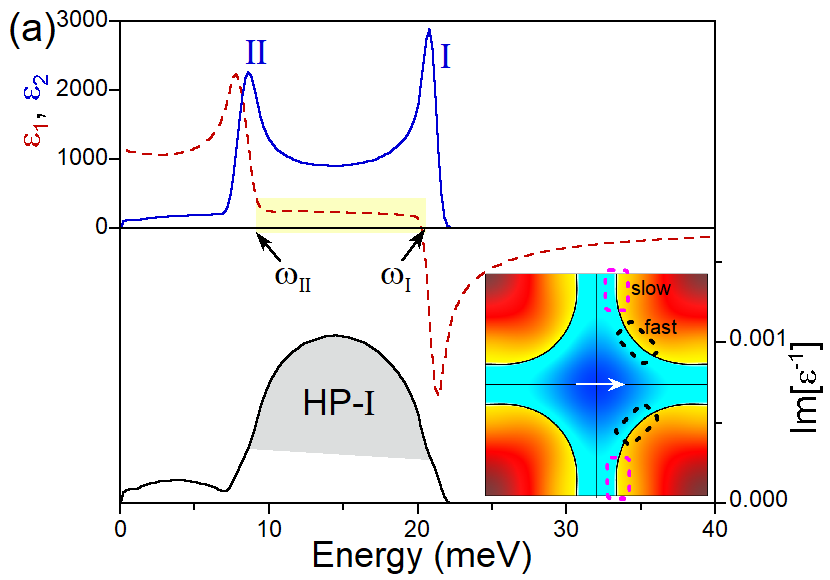}
\includegraphics[width=0.49\linewidth]{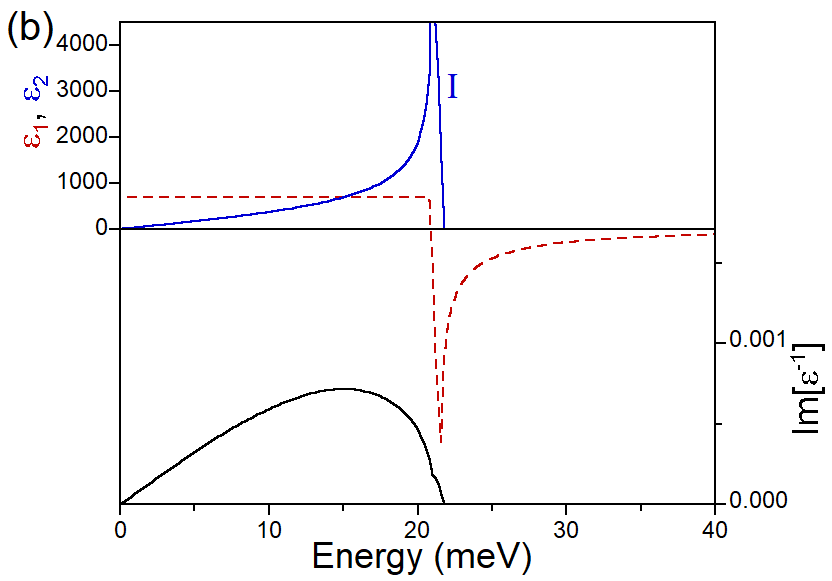}
\includegraphics[width=0.49\linewidth]{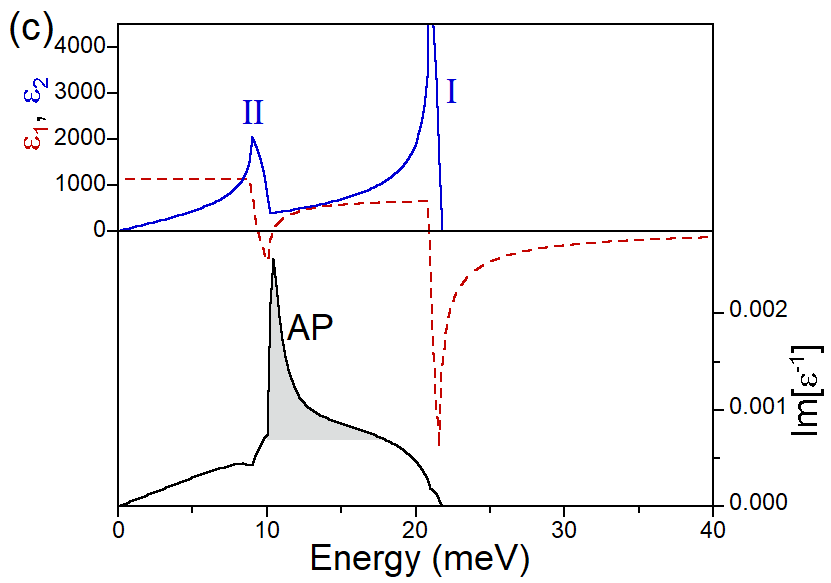}
\includegraphics[width=0.49\linewidth]{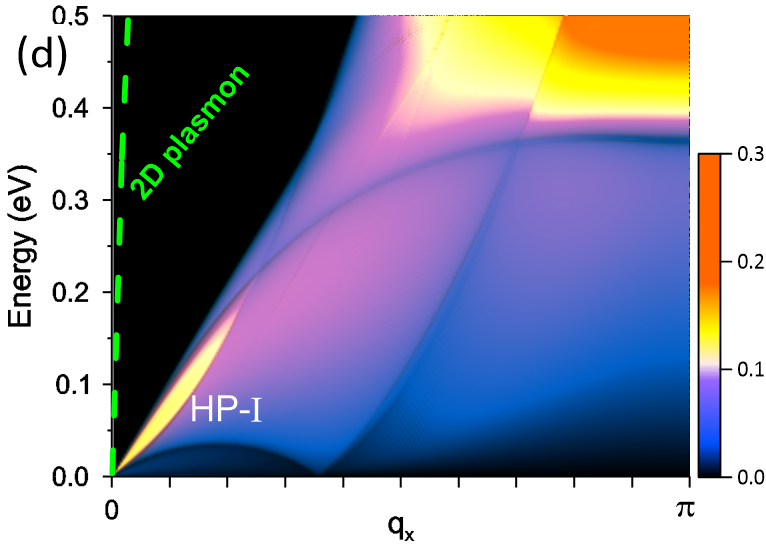}
\caption
{Imaginary (blue solid line) and real (red dashed line) parts of $\epsilon({\bf q},\omega)$  and $-\mbox{Im}[\epsilon^{-1}({\bf q},\omega)]$
(black solid line) evaluated at ${\bf q}=(0.02\pi,0)$ for various models.
(a) TB band dispersion,
(b) one-component electron gas model, and
(c) two-component electron gas model.
In panel a, the yellow rectangle highlights the $\omega_{\rm II}$-$\omega_{\rm I}$ energy interval  where $|\epsilon_1 \ll \epsilon_2$.
Peaks I and II in $\epsilon_2$ are generated by the fast and slow carriers moving in the ${\bf q}$ direction (white arrow in the inset),
respectively.
In the inset the respective regions are schematically highlighted by black and magenta rounded rectangles.
In panel c, peaks I and II are generated by the fast and slow carriers at the Fermi level in two different energy bands. The shaded
regions in the loss functions tentatively represent the spectral weight corresponding to the hyperplasmon of
type I, HP-I, and the acoustic plasmon AP.
(d) The normalized loss function, -Im$[\epsilon^{-1}({\bf q},\omega)]/q$, in the energy-momentum space for ${\bf q}$'s in the $q_x$
direction. Peak HP-I corresponds to the hyperplasmon of type I. The dispersion of a conventional 2D-plasmon is shown by the green dashed line.
}
\label{CUPR_0050_0000}
\end{figure*}

First, we explore the antinodal direction.
The dielectric function $\epsilon({\bf q},\omega)$ and the loss function $-\mbox{Im}[\epsilon^{-1}({\bf q},\omega)]$ evaluated at ${\bf
q}=(0.02\pi,0)$, are presented in Fig. \ref{CUPR_0050_0000}a.
%
The surprising observation is  the energy dependence of the imaginary part of the dielectric function $\epsilon_2({\bf q},\omega)$, which has
a two-peaks structure with a local minimum in between.
To the best of our knowledge for other single-band systems studied so far, only one peak structures have
been documented.
The only exception is given by the case of an extreme band anisotropy studied recently.\cite{musipccp22}
Typical behavior of $\epsilon_2$ in a 2D free-electron gas is illustrated in Fig. \ref{CUPR_0050_0000}b (similar behaviour can be found in the textbooks.\cite{pino66,givi08}) The reason for is that there is only
one band involved (one group of electrons) which leads to a single peak in $\epsilon_2$ only at the top
of the excitation energies. The position of the peak depends on the Fermi velocity of the charge carriers.

In contrast, two well-defined peaks in $\epsilon_2$ due to the intraband transitions can be found in
multiband systems.
The case for two parabolic bands with different Fermi velocities is reported in Fig.\ \ref{CUPR_0050_0000}c.
The presence of two separated peaks in $\epsilon_2$ in Fig.\ \ref{CUPR_0050_0000}c points to two kinds of carriers at the Fermi level
moving with different velocities in the direction of the ${\bf q}$ vector.

The inspection of the Fermi velocity distribution in the Bi-2212 energy band, schematically marked by arrows in the top-right quadrant of
Fig. \ref{2D-band} reveals that the carriers slowly moving in the ${\bf q}=(q_x,0)$ direction reside in the regions highlighted by magenta
rounded rectangles in the inset of Fig. \ref{CUPR_0050_0000}a. The fast carriers moving in the same direction concentrate in the parts
marked by the black rounded rectangles. The dependence of the Fermi-velocity $\upsilon_x$ on momentum $k_x$ is shown at the top of Fig.
\ref{2D-band}.
By symmetry, the same holds also for ${\bf q}$ pointing along the $q_y$ axis.

A comparison of the peak structures of $\epsilon_2$ in Figs. \ref{CUPR_0050_0000}a and \ref{CUPR_0050_0000}c shows a significant
difference in the lower energy peak II between the TB calculations for Bi-2212 and a
two-band free-electron gas system.  While
$\epsilon_2$ in Fig. \ref{CUPR_0050_0000}c grows continuously from zero to the $\omega_{\rm II}$ and then decreases rather
stepwise,
$\epsilon_2$ in Fig. \ref{CUPR_0050_0000}a rises stepwise at $\omega_{\rm II}$ from low values to its maximum and decreases smoothly
with increasing energy.

An unusual feature concerning the real part of the
dielectric function, $\epsilon_1({\bf q},\omega)$, depicted in Fig. \ref{CUPR_0050_0000}a  consists of its flat behavior over an extended energy region between $\omega_{\rm II}$ and $\omega_{\rm I}$ as highlighted by yellow
rectangle. Such a flat behavior of $\epsilon_1$  differs from that observed in the two-band case of Fig.\ \ref{CUPR_0050_0000}c.
In Fig.\ \ref{CUPR_0050_0000}c $\epsilon_1$ crosses zero with a positive slope at a local minimum in $\epsilon_2$. This produces a well-defined peak in the loss function indicating an acoustic plasmon (AP).\cite{picjp56} This mode corresponds to out-of-phase collective
charge oscillations.  Such oscillations were visualized,\cite{siness06} e.g., in the case of an 
acoustic surface plasmon.\cite{sigaepl04,dipon07}

Instead of a plasmon-like sharp peak in the loss function of the two-band free-electron gas model, in $-\mbox{Im}[\epsilon^{-1}]$ in Fig.
\ref{CUPR_0050_0000}a we observe  a rather broad maximum in the energy range of 8-21 meV.
Since $\epsilon_1 \ll \epsilon_2$ in this energy interval, the loss function is well approximated by the
	{\it reciprocal} value of $\epsilon_2$
\begin{equation}
-\mbox{Im}[\epsilon^{-1}]=\epsilon_2/(\epsilon_1^2+\epsilon_2^2) \approx  1/\epsilon_2
\end{equation}
over this entire energy interval.
As a result, the physics of the new collective
excitation is rather different from that of conventional plasmons.
A well-defined plasmon as a conventional bosonic excitation is characterized by a single plasmon energy $\omega_p$ and a lifetime. The latter is inversely
proportional to the
plasmon peak width in the loss function, since the peak energy dependence has the Lorentzian shape.
In contrast, the width of the novel collective mode peak cannot be interpreted as a lifetime. It is just a measure of the energy distance
(confinement) between two peaks in $\epsilon_2$. Moreover, the energy range where this excitation is defined equals fairly the same energy distance.
Comparing the loss functions reported in Figs. \ref{CUPR_0050_0000}a and \ref{CUPR_0050_0000}c one can notice that the spectral
weight of the mode found here is dramatically enhanced over that of a conventional AP. Because of this characteristic, we will refer further
to it as a hyperplasmon of type I (HP-I). This mode corresponds to the out-of-phase collective charge oscillations of states in the
antinodal and nodal directions.

The dispersion of the HP-I peak in the ($q_x,0$) direction
can be traced back in the loss function map shown
in Fig.\ \ref{CUPR_0050_0000}d. This mode has a sound-like dispersion, i.e.,\ its energy vanishes linearly upon ${\bf q}$ approaching
zero. A clear HP-I peak  can be followed up to an energy of about 200~meV. The calculated dispersion of a conventional 2D-plasmon  is shown
by the green dashed line. Its existence in a 2D system and transformation into a 3D-excitation in high-$T_c$ cuprates was discussed in a number of publications\cite{krmoprb88,gryaprb16,gryacp19,nazhprl20,gryaprb20,sihuprb22} since the 1980s.
It is exactly the mode detected in
recent RIXS experiments.\cite{hechn18,liyuqm20,sihuprb22} Some additional features, that we interpret as e-h excitations, can be seen in the
excitation spectrum of Fig.\ \ref{CUPR_0050_0000}d. Noteworthy, the excitation probability of
the e-h pairs in the low-energy region is significantly reduced in comparison with that for
the one-band and two-band free-electron gas systems. The only enhancement  in the loss function at finite momenta and low energies is observed at
$q_x\approx 0.36\pi$. It is related to the band structure nesting at the Fermi level between the flat regions in the vicinity of the
$(0,-\pi)$ and $(0,\pi)$ points of BZ.

\begin{figure*}
\centering 
\includegraphics[width=0.49\linewidth]{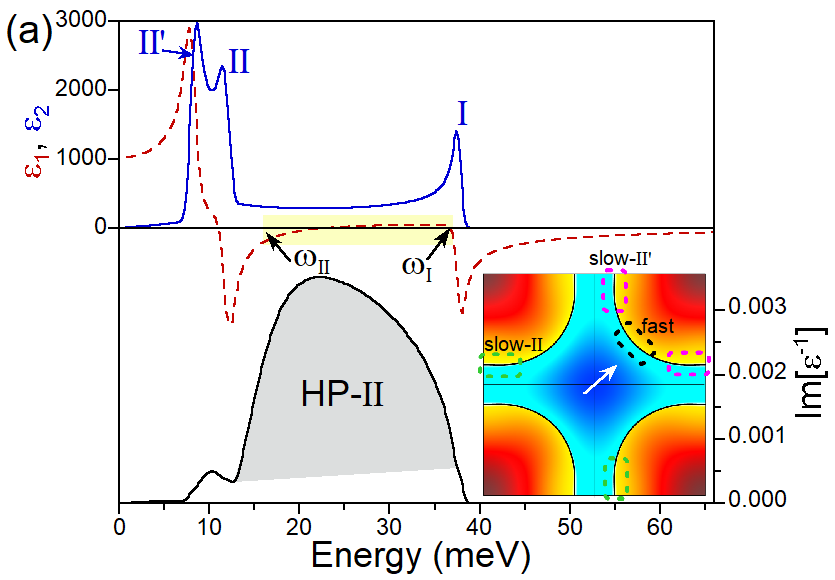}
\includegraphics[width=0.49\linewidth]{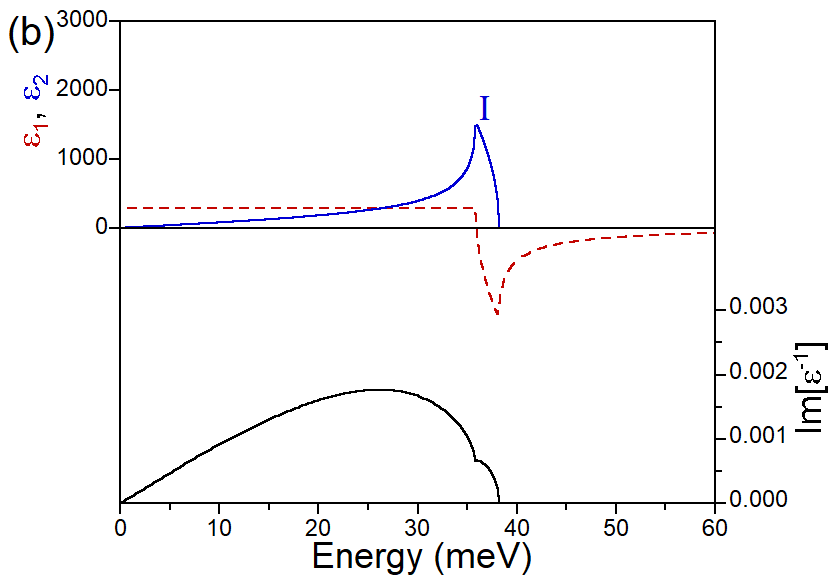}
\includegraphics[width=0.49\linewidth]{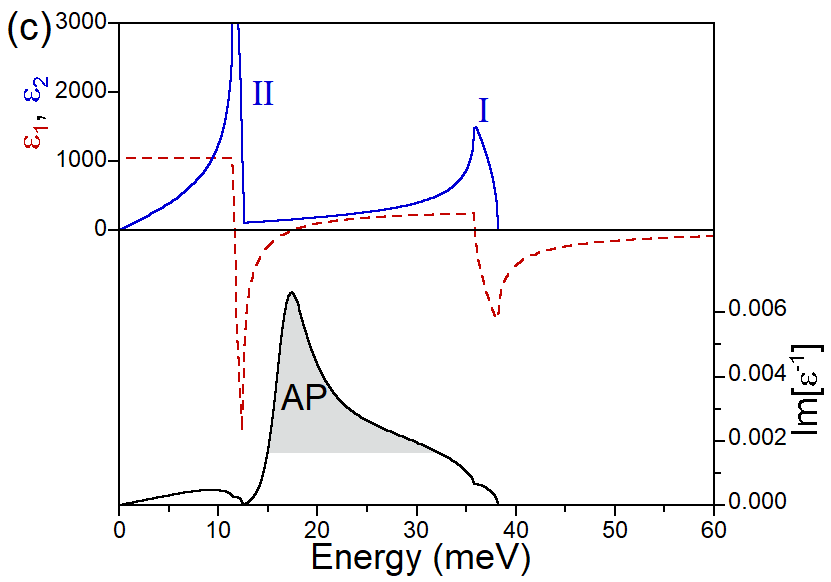}
\includegraphics[width=0.49\linewidth]{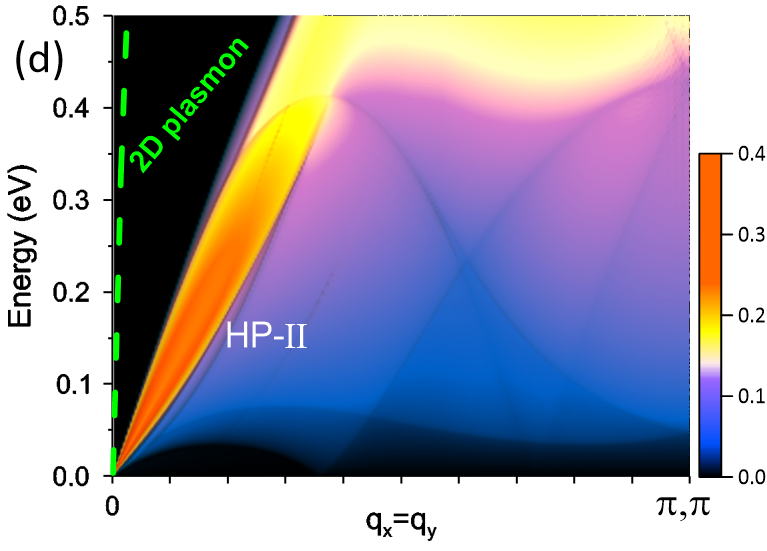}
\caption
{
Imaginary (blue solid line) and real (red dashed line) parts of $\epsilon({\bf q},\omega)$  and $-\mbox{Im}[\epsilon^{-1}({\bf q},\omega)]$
(black solid
line) evaluated at ${\bf q}=(0.02\pi,0.02\pi)$ for various models.
(a) stands for the TB band dispersion,
(b) an one component electron gas model, and
(c) a two-component electron gas model.
In (a) the yellow rectangle highlights the $\omega_{\rm II}$-$\omega_{\rm I}$ energy interval where $|\epsilon_1|\ll \epsilon_2$. The
peaks I, II, and II$'$ in $\epsilon_2$ are generated by the fast and slow carriers moving in the ${\bf q}$-direction (white arrow in the
inset), respectively.
In the inset the respective regions are schematically highlighted by black, green, and magenta rounded rectangles.
In (c) the peaks I and II are generated by the fast and slow carriers at the Fermi level in two different energy bands. The shaded
regions in the loss functions tentatively represent the spectral weight corresponding to the hyperplasmon of type II, HP-II, and
the acoustic plasmon AP.
(d) The normalized loss function, -Im$[\epsilon^{-1}({\bf q},\omega)]/q$, in the energy-momentum space for ${\bf q}$'s in the $q_x=q_y$
direction.
The peak HP-II corresponds to the hyperplasmon of type II. The dispersion of a conventional 2D-plasmon is shown by the green dashed line.
}
\label{CUPR_0050_0050}
\end{figure*}

Figure \ref{CUPR_0050_0050}a shows the dielectric function and loss function calculated at ${\bf q}=(0.02\pi,0.02\pi)$.
For comparison, in Figs. \ref{CUPR_0050_0050}b and \ref{CUPR_0050_0050}c we report the dielectric function and loss function
evaluated in the one- and two-component free-electron gas models at the same momentum ${\bf q}$. One can see that the shapes of the curves are
similar to those in Figs. \ref{CUPR_0050_0000}b and \ref{CUPR_0050_0000}c.
Like in the case of Fig. \ref{CUPR_0050_0000}a, $\epsilon_2$ in Fig. \ref{CUPR_0050_0050}a consists of two essential features. Most of nonzero $\epsilon_2$ concentrates in the energy window between 7 and 40 meV. It is bound by a single peak at the top and a
double-peak structure at the bottom.
The pronounced peak I is generated by the intraband transitions involving the states with maximal Fermi velocities in the region marked in
the inset by a black rounded rectangle. Peak II is produced by "slow" states residing in the regions marked in the inset by green rounded
rectangles. Even slower carriers in the states inside the purple rounded rectangles create a II$'$ peak at lower energies.
The shape of $\epsilon_2$ between  $\omega_{\rm II}$ and $\omega_{\rm I}$ differs from that in the two-band free-electron gas case of Fig.
\ref{CUPR_0050_0050}c.
 All of these modifications result in a qualitatively different behavior of $\epsilon_1$ in the energy region between $\omega_{\rm II}$ and
 $\omega_{\rm I}$. In Fig. \ref{CUPR_0050_0050}a $\epsilon_1$ crosses the zero with a positive slope at an energy around 20 meV.
However, this zero-crossing in $\epsilon_1$ (even accompanied by a local minimum in $\epsilon_2$) does not lead to a well-defined peak
in the loss function at all. Instead, in the loss function, one observes a broad maximum expanded over a substantial energy interval between
$\omega_{\rm II}$ and $\omega_{\rm I}$, similarly to the case shown in Fig. \ref{CUPR_0050_0000}a. Our explanation of such behavior
consists of the behavior of the loss function in this energy region being determined by the dominance of $\epsilon_2$, where the exact shape of
$\epsilon_1$, once its amplitude is small in comparison to $\epsilon_2$, is irrelevant.
Nevertheless, in order to refer to the zero-crossing in $\epsilon_1$, we define the respective collective excitation as a hyperplasmon of
type II, HP-II.
	%
	%
Comparing the spectral weight of the HP-I and HP-II modes, apparently for the latter it is notably larger.

The HP-II dispersion presents a sound-like behavior as seen in Fig. \ref{CUPR_0050_0050}d. Clearly, it is a dominant feature in the
electronic excitation spectrum at these energies. Noteworthy, the increase in the spectral weight of HP-II is accompanied by a pronounced
washing out of the e-h pairs' excitations at energies below about 50 meV.

\begin{figure*}
\centering 
\includegraphics[width=0.48\linewidth]{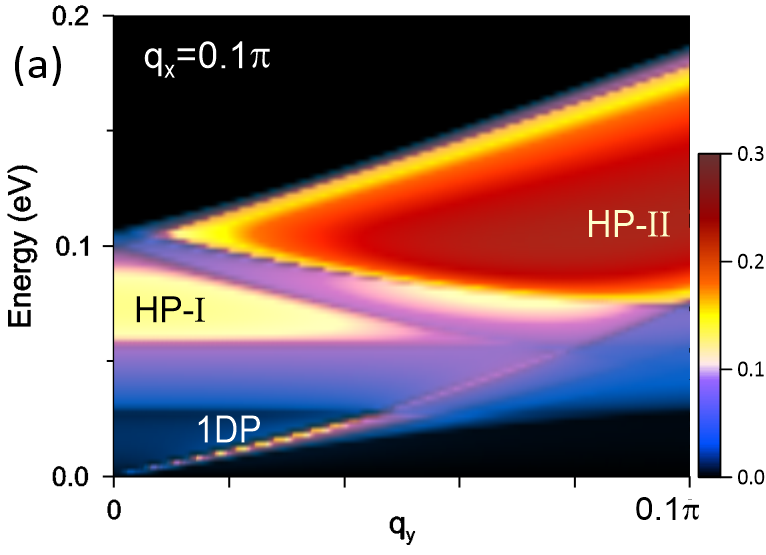}
\includegraphics[width=0.51\linewidth]{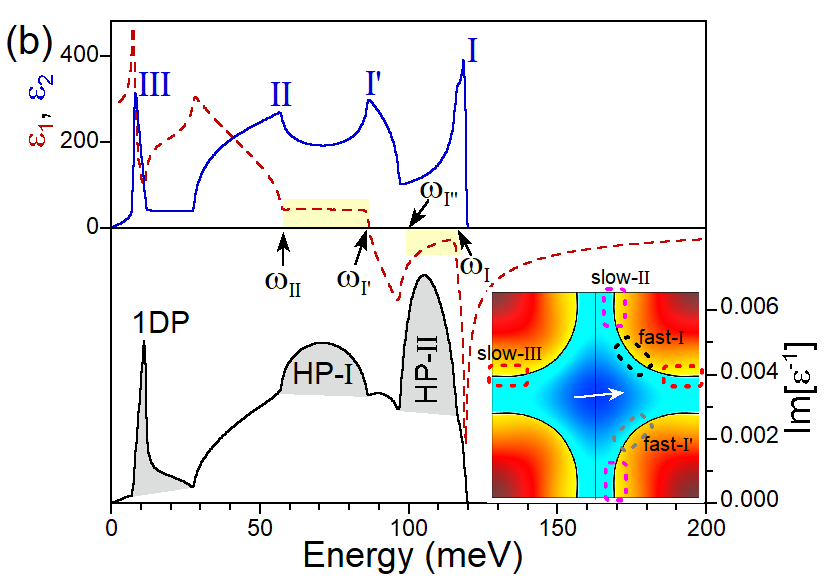}
\caption {
(a) Normalized loss function, -Im$[\epsilon^{-1}({\bf q},\omega)]/q$, in the energy-momentum space for ${\bf q}$'s with $q_x=0.1\pi$ and
$q_y$ varying from 0 to $0.1\pi$. The peaks HP-I, HP-II, and 1DP correspond to the hyperplasmons of the types I and II and the quasi-1D
plasmon, respectively.
(b) Imaginary (blue solid line) and real  (red dashed line) parts of $\epsilon({\bf q},\omega)$ and $-\mbox{Im}[\epsilon^{-1}({\bf q},\omega)]$
(black solid line) evaluated at ${\bf q}=(0.1\pi,0.02\pi)$, obtained with the TB band structure.
The peaks in $\epsilon_2$, labeled I, I$'$, II, and III, are produced by fast and slow carriers
at Fermi level moving in the  ${\bf q}$-direction (white arrow in the inset).
The respective regions with these carriers are marked in the inset by rounded black, gray, magenta, and red rectangles.
The yellow rectangles highlight the $\omega_{\rm II}$-$\omega_{\rm I'}$ and
the $\omega_{\rm I''}$-$\omega_{\rm I}$ energy intervals where
$|\epsilon_1|\ll \epsilon_2$.
 The shaded regions in the loss function represent the spectral weights corresponding to the
HP-I and HP-II hyperplasmons and 1DP.
}
\label{CUPR_0250_0050}
\end{figure*}

A close inspection of the loss function and $\epsilon_2({\bf q},\omega)$ at small energies in Fig. \ref{CUPR_0050_0000}a shows
nonlinear dependence on energy as $\omega^{\alpha}$ with $\alpha <$1 in contrast to those of the free-electron gas models in Figs.
\ref{CUPR_0050_0000}b and \ref{CUPR_0050_0000}c. In order to address this point and trace the dispersion of the HP-I and HP-II
modes, in Fig. \ref{CUPR_0250_0050}a we present the loss function evaluated at ${\bf q}=(0.1\pi,q_y)$ with $q_y$ ranging from 0 to
$0.1\pi$.
One can see that, indeed, the HP-I and the HP-II peaks are different modes and are  not mutually connected.
Upon the departure of $q_y$ from zero,
the intensity of the HP-I peak is quickly reduced and its energy shifts downward.
At $q_y\approx 0.06\pi$ this mode disappears. On the other hand, a notably more intense HP-II peak appears almost immediately upon departure
from $q_y=0$ and quickly gains in intensity. Its energy dispersion is positive.

\begin{figure*}
\centering 
\includegraphics[width=0.49\linewidth]{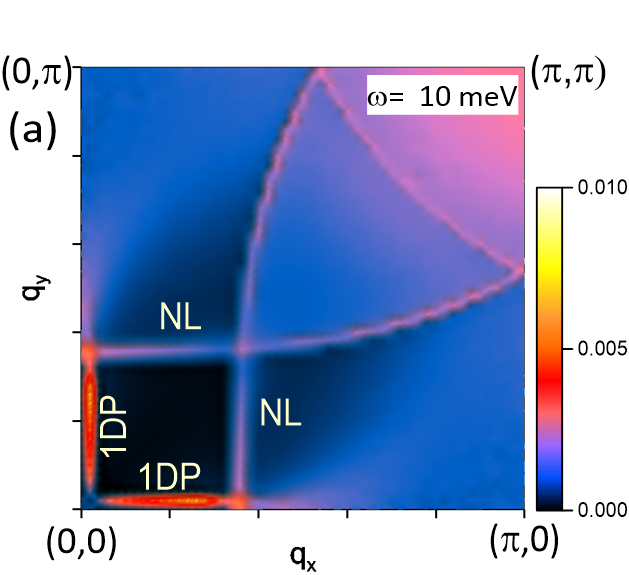}
\includegraphics[width=0.49\linewidth]{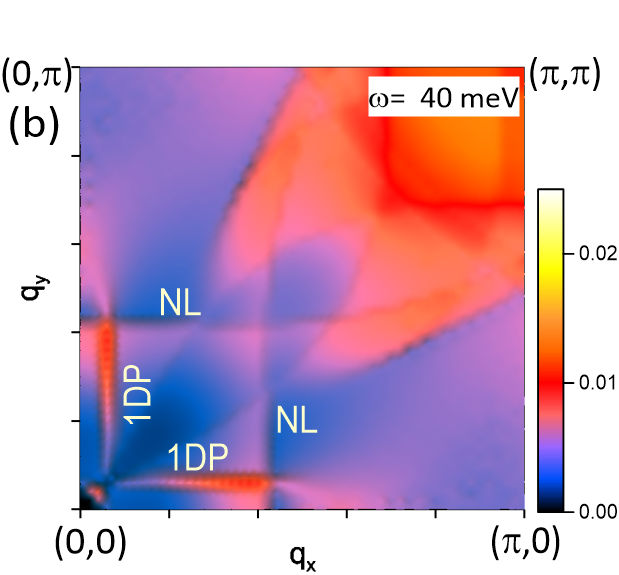}
\includegraphics[width=0.49\linewidth]{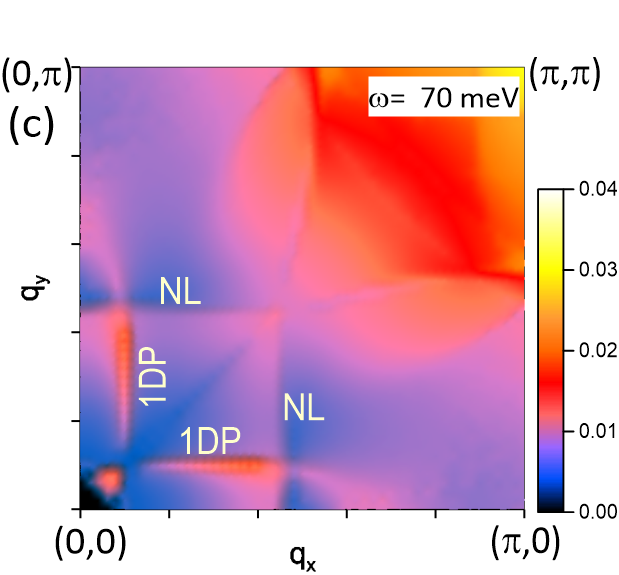}
\includegraphics[width=0.49\linewidth]{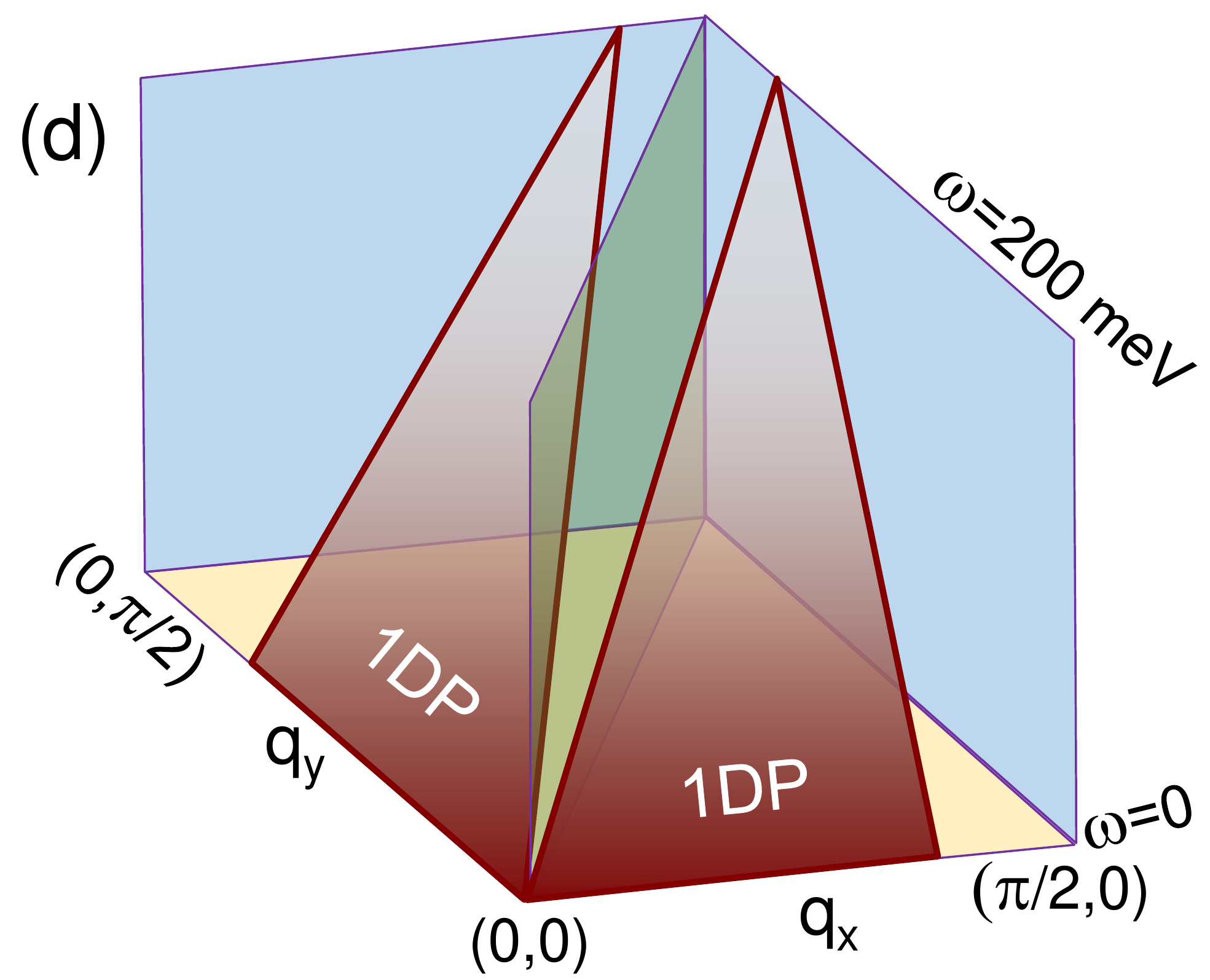}
\caption {Loss function, $-\mbox{Im}[\epsilon^{-1}({\bf q},\omega)]$, versus momentum for (a) $\omega=10$ meV, (b) $\omega=40$ meV,
and (c) $\omega=70$ meV. The peaks corresponding to the quasi-one-dimensional plasmon are marked as 1DP. The features arising from the
Fermi surface nesting are marked as nesting lines (NLs). In (b) and (c) a feature emerging in the left-bottom corner corresponds to the HP-II.
In (d) a 3D pictorial view of the 1DP dispersion is presented.
}
\label{LOSS_w=010-070meV}
\end{figure*}

At the low-energy side of Fig. \ref{CUPR_0250_0050}a one can observe a sharp peak emanating from zero at the $q_y=0$ point.
Apparently it has a sound-like dispersion as a function of $q_y$. However, it occurs at a finite $q_x$ value. Actually, the emergence of
this peak occurs over a ${\bf q}=(q_x,0)$ region with $q_x$  less than 0.36$\pi$  (equivalently, for ${\bf q}=(0,q_y)$ as a function of
$q_y$). It can be seen in Fig. \ref{LOSS_w=010-070meV}a where the loss function distribution map evaluated at $\omega=10$ meV over an
extended momentum region is reported.
The sharp peak, interpreted as a quasi-one-dimensional plasmon (1DP), has a linear shape along the $q_x$ direction. By symmetry, a similar mode
is observed along the $q_y$ direction.
Such flat behavior is a signature of a one-dimensional propagation of such modes. Interestingly, the expansion of the 1DP peaks in the
$(q_x,q_y)$ space is limited by the lines with high intensity produced by the nesting between the flat regions in the Fermi surface. Upon the
$\omega$ increase the flat 1DP peak lines move up (right) with loss of its intensity close to the nodal direction. This is accompanied by the
broadening of the respective peaks as can be appreciated in the loss functions at $\omega$=40 and 70 meV reported in Figs.
\ref{LOSS_w=010-070meV}b and \ref{LOSS_w=010-070meV}c, respectively.
 We estimate that the 1DP peaks can be detected in the loss function up to the energies of about 200 meV.  A pictorial view of its dispersion is presented in Fig. \ref{LOSS_w=010-070meV}d.

 To trace the origin of the 1DP, in Fig. \ref{CUPR_0250_0050}b we report $\epsilon({\bf q},\omega)$ and $-\mbox{Im}[\epsilon^{-1}({\bf
 q},\omega)]$ calculated at ${\bf q}=(0.1\pi,0.02\pi)$, i.e., pointing slightly out of the antinodal direction. Such a small step out of the
 antinodal direction produces dramatic modifications  in $\epsilon_2$.
  A new peak III emerges in $\epsilon_2$ on the low-energy side. It has an origin in the carriers residing in the regions of the Fermi
  surface marked by  red rounded rectangles in the inset of Fig. \ref{CUPR_0250_0050}b.
The possibility of the transitions involving these states arises due to the symmetry breaking.
The sharpness of the III peak can be explained by the flatness of the respective Fermi surface regions. This  results in fairly the same Fermi
velocity of the carriers in any direction.
   The sharp peak in $\epsilon_2$ produces a sudden drop in the real part of the dielectric function  seen at the nearby energy. Such
   behavior of $\epsilon$ results in the appearance of a well-defined peak 1DP in the loss function at a slightly larger energy.
  Therefore, despite the fact that the real part of the dielectric function does not cross zero at the respective energy we interpret this feature as a conventional plasmon.


Additionally, in Fig.\ \ref{CUPR_0250_0050}b one can see that the peak I in $\epsilon_2$ (existing at ${\bf q}=(0.1\pi,0)$) breaks out
into two peaks I and I$'$,
since the group velocities of the states in the regions marked by black and gray rounded rectangles become different.
The shape of peak II maintains almost the same as at ${\bf q}=(0.1\pi,0)$ since the differences in the Fermi velocities in the magenta
regions in the inset of Fig.\ \ref{CUPR_0250_0050}b are small. The emergence of the peak I$'$ in $\epsilon_2$ breaks the region where
$|\epsilon_1|<<\epsilon_2$ into two as highlighted by yellow rectangles. The lower-energy part  maintains its character with a constant value
of $\epsilon_1$, which is a characteristic of HP-I.
In the case of the upper-energy mode, the behavior of the real part of $\epsilon$ is similar to that in the case of HP-II.

Since the dispersion of the energy band in bulk cuprates has almost 2D character,\cite{masaprb05,lenarmp06,scrmp12,kekin15} we expect that
the same collective excitations should exist in the bulk crystals as well. Again, it is related to their independence of the exact expression
of the interparticle potential. Once the band dispersion in the direction perpendicular
to the Cu-O plane is small, its effect on the shape of $\epsilon$ and $-\mbox{Im}[\epsilon^{-1}]$ should also be small.
Consequently, the 2D excitations - HPs  and  1DP - should exist in the rare cases of 3D cuprates\cite{lizhpnas19} as well.

The origins of HPs and 1DP caused by a special shape of $\epsilon_2$ are entirely determined by the shape of the Fermi surface and the velocity
distribution of the carriers there. Once the Fermi surface is similar to that shown in Fig. \ref{2D-band}, HPs and 1DP emerge regardless of
the interparticle interaction that, of course, should be sufficiently strong.

HPs as 2D excitations are excellent candidates to explain the observed
lower-energy peak in the RIXS experimental spectra in the 0-0.4 eV range.\cite{hechn18}
Indeed, the peak ascribed there to a paramagnon, might be a composite excitation consisting of a HP and a paramagnon, although the existence of the latter at small momenta, to the best of our knowledge, was so far not confirmed in the inelastic-neutron scattering experiments.\cite{fokeprb96,hebos02,kekiprl02,hamon04,hipan04,reisprb08,xugunp09,liban10,yuliprb10,fuhijpsj12,chtaprl16} The interrelation
between the HPs and the paramagnon may be  rather intricate and might be a key to understand the hardening or softening of the respective mode
with doping variation in the electron- \cite{wixxprl06,lelenp14,isfunc14} and hole- \cite{lexxnp11,dexxnm13,demiprb17} doped compounds.
We suggest that, at an optimal doping in the small momenta spectra, the measured low-energy bosonic excitation consists of a mixture of the HP
and the paramagnon. Probably the HP
has a larger spectral weight if the Coulomb interaction dominates the spin-flip processes. Upon the momentum increase the paramagnon
gradually gains spectral weight at the expense of the HP
and becomes the
dominating contributor at large momenta.
It would be interesting to scrutinize  the RIXS experimental data\cite{hechn18} related to the measured lower-energy 2D-peak from this
perspective.

It would also be interesting to find other experimental tools sensitive to charge excitations such as electron-energy-loss spectroscopy
(EELS) in reflection mode.
Recent EELS measurements\cite{mihupnas18,humiprx19} demonstrated unusually large spectral weight in the momentum-energy phase space where
the layered plasmon, the HP, and the 1DP can exist.
  Probably new EELS experiments are needed taking into account the existence of several electronic modes. One of the possible ways to realize
  it might consist of the use of cuprate monolayers, like
 those grown recently,\cite{yuman19} in order to  reduce the influence of the substrate and increase the energy separation between the
 2D-plasmon and the modes found here.

Due to a sound-like dispersion, the modes found here cannot be probed directly in the optical experiments. However, in the nanostructures they can become optically active like it is largely exploited in the graphene plasmonics.\cite{jabuprb09,grponp12,gaacsp14,loavacsn14,coganc14,cosiacsn16,mabuoe17,demaaplp18,agbojpcl23} Another possibility might consist of employing optical means in a way similar to that employed for measuring the whole phonon dispersion.\cite{yuleprl91,davinp15,maforpp16,bamaprl17,malijpcl22,cacapss22,lizhjpcl22,repiprb22}

Since the amplitude of the superconducting gap amounts  $\sim 4.5~ k_BT_c \approx 27$~meV  covers the energy region considered here, the e-h
continuum is expected to be suppressed
deep in the superconducting state at low-$T$ outside the nodal regions. As a result, the broad HP features might be become sharper, underlying
the reduced role of the e-h continuum
considered here. Anyway, extending the present approach, the changes of
HPs and 1DP at very low-$T$ in the superconducting state
as well as a more detailed quantitative description of the dielectric function ignored here for the sake of simplicity and clarity are worth
being studied theoretically in near future.
 Corresponding low-$T$ experiments would therefore be of considerable interest and helpful.

In conclusion, we showed that the characteristic electronic structure of Fig. \ref{2D-band} in optimally doped cuprates has a dramatic effect
on the low-energy electronic excitations. The energy region usually considered as being dominated by incoherent electron-hole pairs contains
additionally three kinds of collective charge excitations. A collective mode called here hyperplasmon of type I presents a remarkable
constant-energy region over an extended energy interval, i.e., it is characterized by an uncertainty in energy. It is determined by the difference in the Fermi velocities
 of two groups of states in the flat regions around, for the first group, the $(\pi,0)$ (or the $(0,\pi)$) points in the Brilloun zone and,
 the second one, close to the nodal directions. Such a mode has a maximum present along the antinodal direction.
 A second type of hyperplasmon is realized over a more extended momentum space with a maximal intensity along the nodal direction. This mode
 is more powerful in comparison to the hyperplasmon of type I. The broad peaks in the loss function corresponding to these two modes exhibit
 a non-Lorentzian shape, and their widths cannot be interpreted as a measure of their lifetimes.

Additionally, in the momentum region close to the antinodal direction, we found a third long-lived collective excitation. Its sharp peak in
the loss function has a conventional Lorentzian-type shape and can be used for the lifetime determination. An unusual feature of this
excitation is its dispersion. Its zero-energy limits do not end at a single $(0,0)$ momentum as it happens in the case of a conventional
acoustic plasmon. This mode is soft over an extended momentum range along the antinodal direction.


{\em Acknowledgements}
The authors thank J. Fink and M. Knupfer for their useful critical discussions.
V.M.S. acknowledges financial support by Grant No. PID2019-105488GB-I00 funded by MCIN/AEI/10.13039/501100011033/.
D.E. acknowledges financial support by DFG through the projects 405940956 and 449494427.

\end{document}